\documentclass[pra,showpacs,twocolumn]{revtex4}%
\usepackage{amsmath}
\usepackage{graphicx}
\usepackage{amsfonts}
\usepackage{amssymb}%
\begin{document}
\title{Oscillations of atomic fermions in a one dimensional optical lattice}
\date{\today }
\author{T. A. B. Kennedy}
\affiliation{Dipartimento di Scienze CC,FF,MM, Universit\`{a}
dell'Insubria, via Valleggio 11, 22100 Como, Italia}
 \affiliation{ School of
Physics, Georgia Institute of Technology, Atlanta, Georgia
30332-0430} 

\begin{abstract}
A semiclassical model is used to investigate oscillations of
atomic fermions in a combined magnetic trap and one dimensional
optical lattice potential following axial displacement of the
trap. The oscillations are shown to have a characteristic small
amplitude, damped behavior in the collisionless regime. The
presence of a separatrix in the semiclassical Brillouin zone phase
space is predicted and shown to produce a strongly asymmetric
phase space distribution function.
\end{abstract}
\maketitle

\section{Introduction}
The recent advances in the cooling of atomic Bose and Fermi gases
to degeneracy, and the simplicity with which the gas can be
confined in a periodic optical lattice potential provide a new
point of departure for quantum transport studies of atomic bosons
\cite{Cataliotti,Anderson,Morsch,Denschlag,Greiner} and fermions
\cite{Hofstetter,Rodrigues2,Albus2,LENS}. In particular, neutral
atomic fermions form a nearly ideal gas at nanokelvin
temperatures, and by contrast to conduction electrons in solids,
can traverse the dimensions of the Brillouin zone without
collision.

In a recent experiment a gas of fermionic potassium 40 atoms was
evaporatively cooled in a combined magnetic and optical lattice
potential, leading to a degenerate cloud at a temperature of a
fraction, approximately, one third the Fermi temperature $T_F$ for
the system of $N$ atoms \cite{LENS}. Specifically, the experiment
involved a configuration with a one dimensional axial lattice
potential produced by a laser standing wave field, with tight
magnetic confinement in the transverse plane, but a weaker
magnetic trap in the direction coaxial with the lattice thereby
emphasizing the latter's role. By displacing the center of the
axial magnetic field with respect to the lattice, dipolar
oscillations of the Fermi cloud in the lattice potential were
investigated. The principal observations were that the amplitude
of the oscillations were significantly smaller than the trap
displacement, and were damped over several periods of the motion.

In this paper we show that these observations can be explained by
a simple semiclassical theory of collisionless atomic fermion
motion in the combined magneto-optical potential. This confirms
the qualitative physical picture discussed in ref.\cite{LENS}. The
effects of fermion statistics are felt only in the initial
conditions. A broad distribution of occupied quasi momentum states
with large radial quantum numbers is created by evaporative
cooling on account of the Pauli exclusion principle. In a
semiclassical picture the lattice transport dynamics can in
principle be determined by following the particle trajectories in
a Brillouin zone phase space consisting of atomic position and
quasi momentum. In the problem of interest here, we are able to do
this explicitly in the reduced two dimensional phase space of
axial components of position and quasi momentum. The effects of
radial motion are readily accounted for by an appropriate
summation over radial quantum numbers.

The remainder of this paper is organized as follows. In section II
we develop a theoretical model for the dynamics of the degenerate
cloud in the confined magnetic and optical lattice potentials. In
section III we discuss the semiclassical distribution function on
the axial phase space, and illustrate how the trap displacement
influences the evolution of the distribution function, and the
center of mass motion of the cloud. In section IV we summarize our
conclusions.

\section{Semiclassical atomic phase space trajectories}

The Hamiltonian for single particle motion in a combined optical
lattice and magnetic confining potential may be usefully written
in the following form
\begin{equation}
H = H_0 + H_1 + V(z),
\end{equation}
where $H_0 = (p_x^2+p_y^2)/(2m) + m \omega_r^2(x^2+y^2)/2$, $H_1 =
p_z^2/2m + V_0 sin^2 k_Lz$ and $V(z) = m\omega_a^2 z^2/2$. The
Hamiltonian $H_0$ governs the radial harmonic motion produced by
the magnetic trap, while $V(z)$ is the residual part of the
magnetic confinement in the axial direction. For tight radial
confinement $\omega_r >> \omega_a$. The Hamiltonian $H_1(z) =
H_1(z+a)$ describes one dimensional motion in an optical lattice
with period $a = \lambda_L/2$, where $\lambda_L$ is the optical
wavelength and $k_L = 2 \pi / \lambda_L$ is the wavenumber. The
lattice depth $V_0 = s E_r$ is conventionally parameterized by a
dimensionless multiple s, of the atomic recoil energy $E_r =
\hbar^2 k_L^2 / (2m)$.

We are here interested in the following specific experimental
situation. A single component Fermi gas is cooled to degeneracy in
the combined magnetic and optical lattice potential to well below
the Fermi temperature $T_F$. The axial potential is then rapidly
displaced $V(z) \rightarrow V(z-d)$, and the degenerate gas
performs oscillations in the lattice. These oscillations were
observed by waiting for some time before switching off both
magnetic and trap potentials and imaging the cloud after a period
of ballistic expansion \cite{LENS}.

The one dimensional lattice Hamiltonian $H_1$ has Bloch
eigenfunctions $\psi_{\alpha q}(z)$ and eigenvalues ${\mathcal
E}_{\alpha}(q) = {\mathcal E}_{\alpha}(q + K)$ with $K = (2 \pi/a)
\times \mbox{integer}$, any reciprocal lattice vector. Here $\hbar
q$ is the axial quasi momentum associated with translations $z
\rightarrow z+a$ in the periodic lattice, and $\alpha$ is the
integer axial band index. We employ a semiclassical approach to
treat the transport dynamics of the atomic fermions in the axial
direction. The method is well suited to the description of the
collisionless regime relevant to single component ultracold
fermions where s wave scattering is prohibited by the antisymmetry
of the two-body wave function and p-wave scattering is strongly
suppressed. The energy bands implicitly take into account the
axial lattice potential responsible for large changes in the
physical momentum of the moving particle, but which conserve the
quasi momentum $\hbar q$. We assume, consistent with
ref.~\cite{LENS}, that only the fundamental axial band $\alpha =
1$ is populated at the temperatures of interest, and therefore we
will suppress the axial band index. In order to correctly describe
quantum mechanical motion in the plane transverse to the lattice,
radial quantum numbers will arise later in the theoretical
development.

In the semiclassical theory, motion of an atom is considered to be
restricted to a given energy band, and only external forces other
than the lattice, in this case due to the magnetic trap, can
change its quasi momentum. It follows that the axial motion of a
fiducial atom, in response to the axial displacement $d$ of the
trap, is found by solving the equations of motion
\begin{eqnarray} \nonumber
\hbar \dot q(t) &=& - m \omega_a^2 (z(t)-d) \\
\dot z(t) &=& \frac{1}{\hbar} \frac{\partial {\mathcal
E}(q)}{\partial q}.
\end{eqnarray}
We note that these equations are independent of any radial quantum
numbers associated with motion in the transverse xy plane.

 As a simple example consider motion confined to
near the bottom of the fundamental axial band, but with arbitrary
radial excitation. We may write ${\mathcal E}(q) = \hbar^2 q^2 /
(2 m_{eff}) + \mbox{constant}$, with $m_{eff}$ the associated
effective mass. The equations of motion reduce to
\begin{equation}
\frac{d^2 z(t)}{dt^2} = - \Omega^2 \left( z(t) - d \right),
\end{equation}
describing harmonic oscillations about the displaced position $d$
with renormalized frequency $\Omega = \sqrt{m/m_{eff}} \ \omega_a$
\cite{Javanainen}. The harmonic approximation predicts large
amplitude dipole oscillations but cannot correctly describe the
orbits of occupied fermionic states with relatively large axial
quasi momentum far from the band minimum. On the other hand a Bose
condensate would have a narrow quasi momentum distribution
\cite{Denschlag}. The simple harmonic theory indicates that one
should expect large amplitude dipole oscillations in this case, as
has been observed in experiments \cite{LENS,Burger}. An ideal gas
model with a narrow quasi momentum distribution is, however, not
sufficient to correctly describe superfluid effects associated
with an interacting Bose condensate \cite{Kramer}.

In order to treat non harmonic phase space orbits we note that any
axial energy band must satisfy the periodicity requirement
${\mathcal E}(q) = {\mathcal E}(q+K) $. For example, the axial
energy band
\begin{equation}
{\mathcal E}(q) = \frac{2 (\hbar k_L)^2}{m_{eff} \pi^2}
sin^2\left(\frac{\pi q}{2 k_L} \right).
\end{equation}
behaves like $\hbar^2 q^2 / (2m_{eff})$ for small $q$, but also
satisfies the periodicity requirement with period $2k_L$. Thus all
physically distinct states of a single fermionic atom lie in the
range $-k_L \le q \le k_L$. This energy band has the additional
advantage that it produces pendulum dynamics and the model can be
solved analytically. In this case the equations of motion become
\begin{eqnarray}
\hbar \dot q(t) &=& - m \omega_a^2 (z(t)-d) \\
\dot z(t) &=& \frac{\hbar k_L}{m_{eff} \pi} sin\left(\frac{\pi
q(t)}{k_L} \right).
\end{eqnarray}
Eliminating $z$ and defining the dimensionless quasi momentum $Q
\equiv \pi q / (2 k_L)$ we get the conservation equation
\begin{equation}
\frac{d}{dt} \left( \dot Q(t)^2 + \Omega^2 sin^2 Q(t) \right) = 0.
\end{equation}
The first integral may be written $\dot Q(t)^2 + \Omega^2 sin^2
Q(t) = \Omega^2 M$ where $M$ is fixed for a given orbit. The
solution is given by
\begin{equation}
sinQ(t) = \sqrt{M} sn\left(\pm\Omega t +\phi_M | M \right)
\end{equation}
where the initial phase
\begin{equation}
\phi_M = u(sinQ(0)/ \sqrt{M} |M)
\end{equation}
is given in terms of the Legendre elliptic function of the first
kind $u(x|m)$. Using the identities $sn(x|M) = sn(\sqrt M
x|1/M)/\sqrt{M}$ and $\sqrt{M} u(sin\phi|M) = u(\sqrt{M} sin\phi |
1/M)$, we can also write
\begin{equation}
sinQ(t) =  sn\left(\pm \sqrt{M}\Omega t +\phi_{1/M} |1/ M \right)
\end{equation}
where the initial phase $\phi_{1/M} = \phi_M  \sqrt{M}$ is given
by
\begin{equation}
\phi_{1/M} = u(sinQ(0)|1/M).
\end{equation}
We present these results in terms of elliptic functions in which
the amplitude argument, either $M$ or $1/M$, is in the range
$[0,1]$. This is useful for the comparison of orbits with $M < 1$
and $1/M < 1$. As we discuss further below, these orbits are
analogous to the oscillating and rotating orbits, respectively, of
a mechanical pendulum.

Differentiation of the result for $sinQ(t)$ with respect to time
yields
\begin{equation}
\dot Q(t) = \pm\Omega \sqrt{M} cn( \pm \Omega t+\phi_M|M),
\end{equation}
which in turn gives the atomic trajectories
\begin{eqnarray}
z(t)- d = \mp \ell \sqrt{M} cn(\pm\Omega t+ \phi_M|M).
\end{eqnarray}
or, using $cn(x|M) \equiv dn(\sqrt{M}x|1/M)$
\begin{eqnarray} \nonumber
z(t)- d = \mp \ell \sqrt{M} dn(\pm \sqrt{M}\Omega t+
\phi_{1/M}|1/M).
\end{eqnarray}
The value of $M$ depends on the trap displacement and satisfies
the ``energy" conservation condition,
\begin{equation} \label{orbits}
(z(0)-d)^2 + \ell^2 sin^2Q(0) = \ell^2 M.
\end{equation}
where we define the length scale
\begin{equation}
\ell  = \frac{2}{\pi}\sqrt{\frac{m}{m_{eff}}} \left( \frac {\hbar
k_L}{m \omega_a} \right),
\end{equation}
such that ${\mathcal E}(q) = e_0 sin^2Q$ with $e_0 = m\omega_a^2
\ell^2/2$. For $M \le 1$, the orbits in the $z-q$ phase space are
closed, whereas for $M > 1$ they are open. The separatrix orbit $M
= 1$ has homoclinic fixed points on the Brillouin zone boundary,
analogous to the infinite period motion of a physical pendulum
between vertically upwards initial and final positions, $sinQ(t) =
tanh(\pm\Omega t +\phi_1)$ and $(z(t)-d)/\ell = \mp sech(\pm\Omega
t +\phi_1)$. The qualitatively different character of these orbits
is responsible for a non-trivial dependence of the dipole
oscillations on the initial trap displacement. When the trap is
displaced $z \rightarrow z-d$, the peak of the phase space
distribution function remains at $z=0$, but the separatrix is
given by Eq.(\ref{orbits}) with $M=1$, and is centered in the
region $z>0$. Subsequently the atoms lying inside the separatrix
will perform closed orbits, while those left outside remain
outside, and perform small amplitude spatial oscillations as they
traverse the Brillouin zone. The relative importance of the two
sets of orbits in the overall motion of the cloud is determined by
the fraction of atoms initially inside and outside the separatrix.
In turn, this is governed by the magnitude of the initial
displacement $d$ relative to $\ell$. As we will illustrate in the
following section, the axial distribution function can develop
strong asymmetry for large trap displacements, due to the
separatrix structure of the Brillouin zone phase space.

\section{Dynamics of the axial atomic distribution function}
 The initial thermal distribution of atomic fermions produced
by evaporative cooling in the combined magnetic trap and optical
lattice potentials, can be written in terms of a semiclassical
distribution function for motion in the axial direction with the
radial harmonic oscillator quantum numbers $n_x$ and $n_y$. Since
the radial potential is assumed to be azimuthally symmetric, only
the combination $n_x+n_y$ arises. Recall that we assume the
temperature is sufficiently low that only the lowest axial band is
populated \cite{LENS}. We are therefore left to consider a set of
axial distribution functions with different radial quantum
numbers. Specifically, the initial axial distribution function is
given by $f^{(0)}(z,q) = \sum_{n_r = 0}^{\infty}
f^{(0)}_{n_r}(z,q)$, where
\begin{equation}
f^{(0)}_{n_r}(z,q) = (n_r+1)\left[ e^{\beta\left( n_r \hbar
\omega_r + {\mathcal E}(q) + V(z) - \mu \right)} + 1 \right]^{-1},
\end{equation}
with $\beta = 1/(k_B T)$, $n_r =  0,1,2,...$ the radial quantum
number and $\mu$ the chemical potential. Note that, due to the
lattice, this is a function of $z$ and quasi momentum $\hbar q$,
as opposed to physical momentum. The factor $(n_r+1)$ is the
degeneracy of the radial band with energy $\hbar \omega_r n_r$
above the ground state $n_r = 0$. The distribution function is
assumed not to change if the axial trap is displaced rapidly. The
subsequent collisionless motion of the cloud causes the
distribution to change according to the Boltzmann equation
\begin{equation}
\frac{\partial f_{n_r}}{\partial t} + \frac{\partial {\mathcal
E}(q)}{\partial \hbar q}\cdot \frac{ \partial f_{n_r}}{\partial z}
-m \omega_a^2 (z-d) \frac{ \partial f_{n_r}}{\partial q} = 0,
\end{equation}
with solution
\begin{equation}
f_{n_r}(z,q,t) = f^{(0)}_{n_r}(z(t),q(t)).
\end{equation}
Using the previous analysis above we have that, for $M < 1$,
\begin{eqnarray}
sinQ(t) &=& \frac{ sinQ(0)cn\Omega t dn\Omega t -
\frac{(z(0)-d)}{\ell} cosQ(0) dn\Omega t}{1 - sin^2Q(0)sn^2 \Omega
t} \\ \nonumber
z(t)-d &=& \frac{(z(0)-d)cn \Omega t + \ell
sinQ(0)cosQ(0)sn\Omega t dn \Omega t }{1 - sin^2Q(0)sn^2\Omega t}
\end{eqnarray}
where we note that $dn(\phi_M|M) = cosQ(0) \ge 0$ in the first
Brillouin zone, and $sn \Omega t \equiv sn(\Omega t|M)$; similarly
for cn and dn. For $M>1$, the following expressions hold
\begin{eqnarray}
sinQ(t) &=& \frac{ sinQ(0)cn\omega t dn\omega t -
\frac{(z(0)-d)}{\ell\sqrt{M}} cosQ(0) sn\omega t}{1 -
sin^2Q(0)sn^2 \omega t/M} \\ \nonumber z(t)-d &=& \frac{(z(0)-d)dn
\omega t + \frac{\ell}{\sqrt{M}} sinQ(0)cosQ(0)sn\omega t cn
\omega t }{1 - sin^2Q(0)sn^2\omega t/M}
\end{eqnarray}
where we note $cn(\phi_{1/M}|1/M) = cosQ(0)$ and $sn \omega t
\equiv sn(\sqrt{M} \Omega t|1/M)$, etc.

 The axial number density of fermions in radial band $n_r$
at time $t$, is given by integrating over the quasi momentum of a
single Brillouin zone
\begin{equation}
n_{n_r}(z,t) = \int_{-k_L}^{k_L} \frac{dq} {2\pi} f_{n_r}(z,q,t),
\end{equation}
and the total axial column density $n(z,t) = \sum_{n_r =
0}^{\infty} n_{n_r}(z,t)$. Similarly, the distribution of quasi
momentum can be computed by summation of the distribution
functions for the individual radial energy bands, $n_{n_r}(q,t) =
\int dz f_{n_r}(z,q,t)$.
\begin{figure}[ptb1]
\begin{center}
\includegraphics{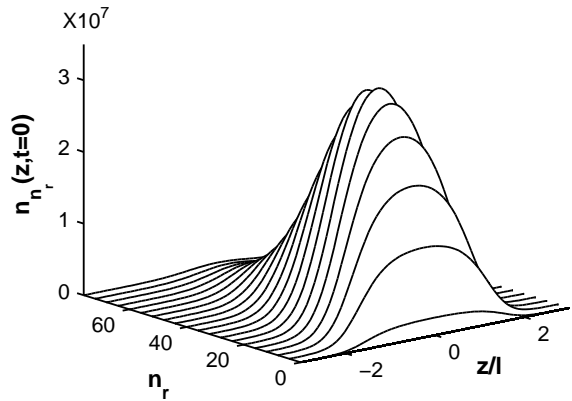}
\end{center}
\caption{The initial atomic density is shown as a function of the
radial quantum number $n_r$. The total axial density $n(z,0)$ is found
 by summation over $n_r$. Parameters are given in the body of the text.}%
\label{n1Fig}%
\end{figure}
\begin{figure}[ptb2]
\begin{center}
\includegraphics[height=6.3cm,width=8.4cm]{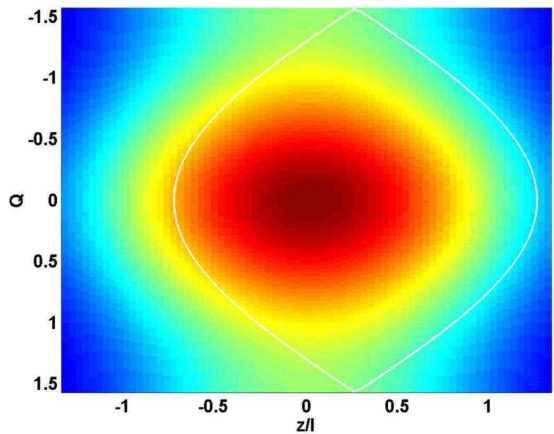}
\end{center}
\caption{The initial axial atomic distribution function in the z-Q
phase space for $d=15\mu$m. The oval shaped contour is the
separatrix across which particles may not cross during the
collisionless dynamics. Other parameters are the same as in
Fig.~\ref{n1Fig}.
 }%
\label{n2Fig}%
\end{figure}
\begin{figure}[ptb3]
\begin{center}
\includegraphics[height=6.3cm,width=8.4cm]{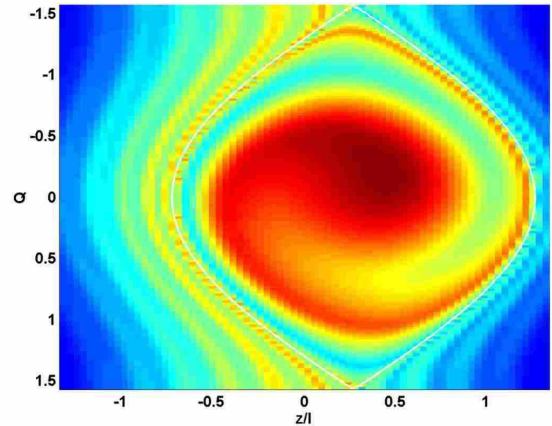}
\end{center}
\caption{The axial atomic distribution function at $\Omega t =15$,
in the z-Q phase space for $d=15\mu$m. The region of high density
lies inside the separatrix, but has been stretched out over the
region. Other parameters are the same as in Fig.~\ref{n1Fig}.
}%
\label{n3Fig}%
\end{figure}
\begin{figure}[ptb4]
\begin{center}
\includegraphics[height=6.3cm,width=8.4cm]{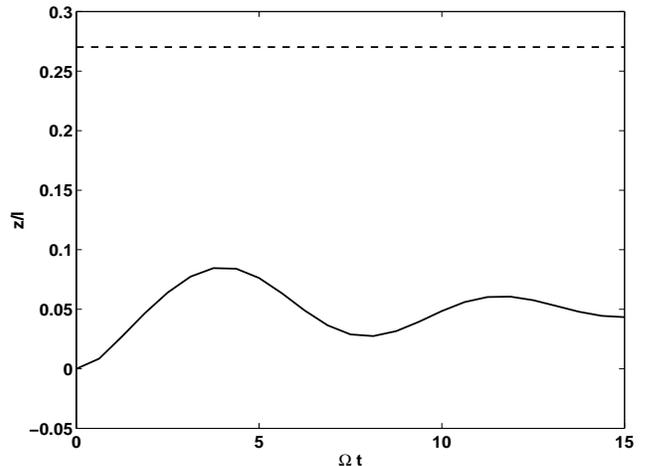}
\end{center}
\caption{Small amplitude damped oscillation of the axial component
of the cloud center of mass near the undisplaced center of the
cloud $z=0$. The position of the displaced trap center is shown by
the dashed line. Other parameters are the same as
in Fig.~\ref{n1Fig}.}%
\label{n4Fig}%
\end{figure}

\begin{figure}[ptb5]
\begin{center}
\includegraphics[height=6.3cm,width=8.4cm]{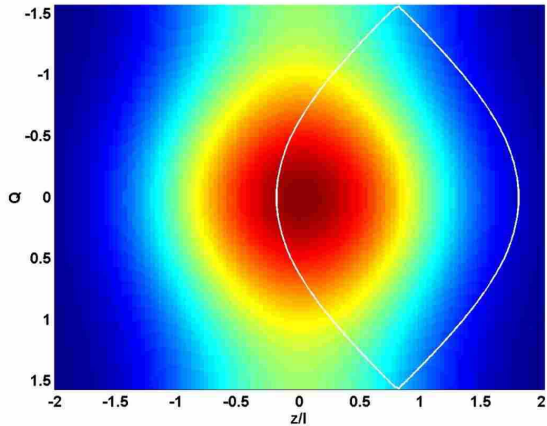}
\end{center}
\caption{The initial atomic distribution function in the z-Q phase
space for $d=45\mu$m. The region of high
density traverses the phase space separatrix. Other parameters are the same
as in Fig.~\ref{n1Fig}.}%
\label{n5Fig}%
\end{figure}

\begin{figure}[ptb6]
\begin{center}
\includegraphics[height=6.25cm,width=8.4cm]{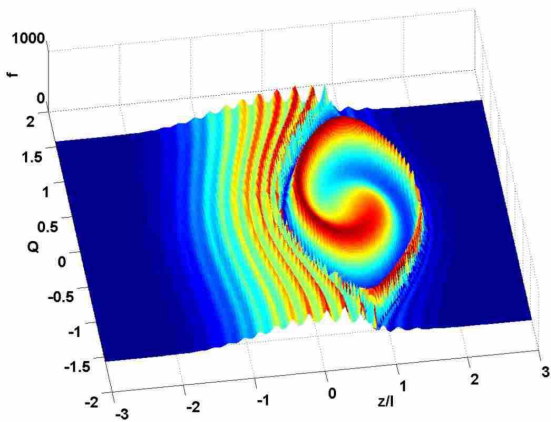}
\end{center}
\caption{The atomic distribution function in the z-Q phase space
at $\Omega t_f =15$ for $d=45\mu$m. The complicated structure
reflects the open and closed orbits in the first Brillouin zone,
which lead to a strong asymmetry. Other parameters are the same
as in Fig.~\ref{n1Fig}.}%
\label{n6Fig}%
\end{figure}

In Fig.1, we show the initial spatial density as a function of
radial quantum number $n_r$ and $z$ at a temperature $T \approx
0.3 T_F$. We have taken the experimental parameters from
ref.~\cite{LENS}, except that we use the analytical model energy
band discussed in the last section rather than the numerical
result appropriate to an optical standing wave. Hence our results
are not intended for quantitative comparison with the experiment.
However, since the qualitative features of the model energy band
are the same as those shown in ref.~\cite{LENS}, its predictions
should be in qualitative agreement with the experimental
observations. The parameters employed in the figures are given
here mainly in units of temperature, therefore energies are
recovered by multiplying by Boltzmann's constant. We consider
potassium-40 atoms cooled in a magnetic trap with radial and axial
frequencies $\omega_r = 15.2 \mbox{nK}$ and $\omega_a = 1.15
\mbox{nK}$, respectively. The wavelength of the incident light is
taken to be $\lambda_L = 754$nm. The chemical potential $\mu = 430
\mbox{nK}$, and $\beta = 6.98 \mbox{(MK)}^{-1}$. For our model
axial energy band, assuming $m_{eff} = m$ for simplicity, this
gives the length scale $\ell \approx 55.5 \mu\mbox{m}$ and band
depth $e_0 \approx 170$nK. A density plot of the initial
distribution function is shown in Fig.2, along with the separatrix
trajectory in the phase space following trap displacement. Orbits
inside and outside the separatrix follow its contour precisely as
occurs in the case of the mechanical pendulum. In this example we
assume a trap displacement $d=15 \mu\mbox{m}$ much less than
$\ell$, so that the bulk of the atomic distribution lies inside
the separatrix, and therefore most though not all particles will
undergo bound orbits in the first Brillouin zone.

The dynamics of the particle distribution in each radial band can
be calculated separately, as explained above, and then summed over
the radial quantum number $n_r$ to get the total axial
distribution function. However, the latter summation is
unnecessary for a qualitative understanding of the dynamics. Each
radial band distribution function has the same dynamics, because
the phase space trajectories $(z(t), q(t))$ are independent of the
radial quantum number $n_r$. In particular the damped motion of
the cloud center of mass is evident in the distribution function
$f_{n_r}(z,q,t)$ dynamics, for a fixed radial quantum number
$n_r$. In Fig.\ref{n3Fig} we show the total axial distribution
which evolves from the initial distribution shown in
Fig.\ref{n2Fig} at time $\Omega t_f = 15$. During this time, the
center of mass of each radial band density undergoes a similar
damped dipole oscillation leading to the center of mass dynamics
for the cloud illustrated in Fig.\ref{n4Fig}. We note that the
amplitude of the oscillation is small compared to the displacement
of the trap center, in qualitative agreement with the observations
of ref.\cite{LENS}. The damping is caused by the stretching of the
distribution function along the phase space trajectories, not by
collisions, dissipation or the influence of different radial
orbits. It is straightforward to check, and is indeed obvious in
the limit of a very narrow distribution in $(z,Q)$ phase space,
that the damping is correspondingly reduced. However, a Fermi
distribution must necessarily have a broad distribution of $Q$
states on account of the Pauli exclusion principle. Damping is
therefore a feature of collisionless Fermi transport dynamics. A
narrow quasi momentum distribution is predicted also by the
Gross-Pitaevski dynamics of a superfluid Bose condensate in a
lattice, and in this case undamped oscillations have been observed
\cite{Burger,LENS}.

The non harmonic orbits of large axial quasi momentum which are
necessarily populated in a Fermi gas cause the stretching of the
distribution function both inside and outside the separatrix. As
noted above, each radial band executes a similar damped motio in
phase with the others. In Fig.\ref{n5Fig} and Fig.\ref{n6Fig} we
illustrate the dynamics of the axial distribution function for a
larger trap displacement $d=45 \mu \mbox{m}$, a value which is
comparable to the separatrix distance $\ell \approx 55 \mu
\mbox{m}$. In this case a fraction approaching one half of the
particles now follow open orbits in the first Brillouin zone. The
total distribution function develops strong whorls inside the
separatrix and spatial oscillations occur in the exterior. The
basic left-right asymmetry survives the integration over the quasi
momentum required to determine the axial spatial density, but the
center of mass motion is qualitatively similar to that illustrated
in Fig.\ref{n4Fig}. Damped oscillations of the Fermi gas alone
cannot be viewed as strong evidence of a separatrix structure in
the Brillouin zone phase space, since both closed and open orbits
produce small amplitude center of mass oscillations. However, the
presence of the separatrix could well be observable experimentally
if the amplitude of center of mass oscillations were
systematically studied as a function of trap displacement $d$.
Alternatively, direct measurements of the axial distribution
function or spatial density would be most useful.

\section{Conclusion}
We have shown that a simple semiclassical theory of the axial
motion of a gas of atomic fermions, without any collisional
interactions, produces transport dynamics in qualitative agreement
with recent experimental observations \cite{LENS}. The gas
executes damped oscillations with an amplitude small compared to
the axial magnetic trap displacement. We also point out the
presence of a separatrix in the Brillouin zone phase space which
produces strong asymmetries in the axial phase space distribution
and axial density functions. It would be interesting to
investigate this structure further experimentally, for example, by
looking for non-monotonic behavior of the transport properties as
a function of the initial trap displacement $d$.
\begin{acknowledgments}
We thank M. Inguscio and G. Roati for interactions which led to
the present work and S.D. Jenkins for discussions. We also
acknowledge the support of a MUIR grant and the hospitality of the
Universit\`{a} dell'Insubria where this work was carried out.
\end{acknowledgments}

\bibliography{ferm}
\end{document}